\documentstyle[12pt]{article}
\newtheorem{Proposition}{Proposition}
\newtheorem{Lemma}{Lemma}

\begin{document}

\title{Noncommutative Unification of General Relativity and
Quantum Mechanics.}

\author{Michael Heller\thanks{Correspondence address: ul.
Powsta\'nc\'ow Warszawy 13/94, 33-110 Tarn\'ow, Poland. E-mail:
mheller@wsd.tarnow.pl} \\
Vatican Observatory, V-00120 Vatican City State
\and Leszek Pysiak \\
Department of Mathematics and Information Science, \\ Warsaw
University of Technology \\
Plac Politechniki 1, 00-661 Warsaw, Poland
\and Wies{\l}aw Sasin \\
Department of Mathematics and Information Science, \\ Warsaw
University of 
Technology \\
Plac Politechniki 1, 00-661 Warsaw, Poland}
\date{\today}
\maketitle

\begin{abstract}
In Gen. Rel. Grav. ({\bf 36}, 111-126 (2004); in press,
gr-qc/0410010) we have proposed a model unifying general
relativity and quantum mechanics based on a noncommutative
geometry. This geometry was developed in terms of a
noncommutative algebra ${\cal A}$ defined on a transformation
groupoid $\Gamma $ given by the action of a group $G$ on a space
$E$.  Owing to the fact that $G$ was assumed to be  finite it
was possible to compute the model in full details. In the
present paper we develop the model in the case when $G$ is a
noncompact group. It turns out that also in this case the model
works well.  The case is important since to obtain physical
effects predicted by the model we should assume thet $G$ is a
Lorentz group or some of its representations. We show that the
generalized Einstein equation of the model has the form of the
eigenvalue equation for the generalized Ricci operator, and all
relevant operators in the quantum sector of the model are random
operators; we study their dynamics. We also show that the model
correctly reproduces general relativity and the usual quantum
mechanics. It is interesting that the latter is recovered by
performing the measurement of any observable. In the act of such
a measurement the model ``collapses'' to the usual quantum
mechanics. 
\end{abstract}

\section{Introduction}
Noncommutative geometry plays an increasingly important role in
the present search for quantum gravity (from a host of papers
let as quote at least a few
\cite{ChFF,ChFG,ChamCon,CZ,Con96,MadMour,Mjd,Sit}). It has also
recently been recognized that it is a useful tool in superstring
theory (the classical paper is \cite{SeibWitt}, and a book
\cite{Kar}). In a series of papers (\cite{Towards,HSL,Red}), we
have proposed our own approach to the unification of general
relativity and quantum mechanics based on noncommutative
geometry. Our starting point is the standard method of changing
a differential manifold (space-time) $M$  into a noncommutative
space \cite[p. 86]{Connes}. It is done by implementing the
following steps: (1) we represent $M$ as a quotient space $N/R$
where $N$ is a suitable space and $R$ a suitable equivalence
relation; (2) then we change from $N/R$ to a suitably organized
subset ${\cal R}$ of $N \times N$; we call this the ``pairing
process''; (3) we define a suitable algebra on ${\cal R}$; and
finally, (4) we extract information about $N/R$ from this
algebra.
\par
We implement this strategy as follows. Let $M$ be a space-time
manifold. The natural way to present $M$ as a quotient space is
with the help of the frame bundle over $M$. Let $\pi_M: E
\rightarrow M$ be the frame bundle with the structure group $G$,
then $M = E/G$. To perform the "pairing process" let us notice
that the group $G$ acts (to the right) on $E$ (along the
fibres), $E \times G \rightarrow E$. We can equip $E \times G$
with the groupoid structure. This groupoid is called a {\it
transformation groupoid\/}, and will be denoted by $\Gamma $
(its description is given in Sect. 2).  Now, we define a
(noncommutative) compactly supported, smooth, complex valued
algebra ${\cal A}$ on $\Gamma $ with convolution as
multiplication. Then we construct, in terms of this algebra, the
(noncommutative) geometry of the groupoid $\Gamma $ which is a
generalization of the usual space-time geometry (on $M$). The
regular reprezentation of the algebra ${\cal A}$ on a bundle of
Hilbert spaces gives us the ``quantum sector'' of the model.
\par
To smooth out some inaccuracies and avoid conceptual traps in
which our prevous work was involved we have tested the method on
a simpler model in which the group $G$ was finite
(\cite{Finite,Random,Obs}). It has turned out that this
simplified model works well. Let us notice, however, that if a
finite group $G$ acts freely on a space $ E$ then $G$ must be a
cyclic group isomorphic with ${\bf Z}^n$ where $ n=|G|$. Indeed,
for $G\ni g\neq e$ the set $\{gp,g^2p,\ldots ,g^np\}$ is
bijective with $ G$ and, as it can be easily seen, $g^n=e$.
However, we should notice that the fact that the group $ G$ is
Abelian does not entail the commutativity of the groupoid
algebra $ {\cal A}$.  Therefore, our model with a finite group
$G$ could serve well as an ``exercise model'', but to have a
more physically realistic approach we must change to an infinite
group $G$. This is exactly what we do in the present paper.
Throughout this paper it is assumed that $G$ is a noncompact
group. This is an important case since to obtain physical
effects predicted by our model we should assume that $G$ is a
Lorentz group or some of its representations.
\par
In Sect. 2, we briefly present the groupoid $\Gamma $ and its
algebra ${\cal A}$, and we establish notation. The geometry of
the groupoid $\Gamma $ is based on the module of derivations of
the algebra ${\cal A}$. In Sect. 3, we study the structure of
this module and, in Sects. 4 and 5, we develop the differential
geometry of the groupoid $\Gamma $. This enables us to
formulate, in Sect.  6, generalized Einstein's equation. It
turns out that it has the form of the eigenvalue equation for
the generalized Ricci operator. We also show that the standard
space-time geometry is obtained by suitably ``averaging''
elements of ${\cal A}$.  In Sect. 7, we study the quantum sector
of the model, and show that all relevant operators are random
operators. We also investigate their generalized dynamics. The
transition from our model to the usual quantum mechanics is
presented in Sect.  8. Interestingly, it is the act of
measurement of any observable that reduces our model to the
usual quantum mechanics. We thus can say that from the
perspective of our model quantum mechanics is but a theory of
making measurements.
\par
The present paper focuses on mathematical aspects of the
proposed model; its physical aspects will fuller be discussed in
a forthcoming paper.

\section{Preliminaries}
Let $\Gamma =E\times G$, where $E$ is the frame bundle over
space-time $M$ with the structural group $G$, such that $
G$ is
a noncompact semisimple Lie group\footnote{Let us
notice that the Lorentz group is noncompact and semisimple.}
acting on $E$, be a transformation groupoid, and ${\cal A}
=C_c^{\infty}(\Gamma ,{\bf C})$ the noncommutative algebra of
smooth, compactly supported, complex valued functions on
$\Gamma$ with convolution as multiplication. Let further
$\gamma_1=(p_1,g_1)$, $\gamma_2=(p_2,g_2)\in\Gamma$, and $
p_2=p_1g$. We assume the convention
$\gamma_1\circ\gamma_2=(p_1,g_1g_2)$, and consequently
\[(f_1*f_2)(\gamma )=\int_{\Gamma_{d(\gamma )}}f_1(\gamma_
1)f_2(\gamma_1^{-1}\gamma )d\gamma_1\]
for $f_1,f_2\in {\cal A}$, where $d(\gamma )=d(p,g)=p$.
\par
Let us notice that the center of the algebra ${\cal A}$
vanishes, ${\cal Z}({\cal A})=\{0\}$, but ${\cal A}$ is a
module over $Z=\pi_M^{*}(C^{\infty}$(M)) (here $\pi_M:E
\rightarrow M$ is the bundle projection). Functions of $
Z$, in general, are not compactly supported. However, they do
act on ${\cal A}$ in the following way: $\alpha :Z\times {\cal
A} \rightarrow {\cal A}$ by
\[\alpha (f, a)(p,q)=f(p)a(p,g),\]
$f\in Z,\,a\in {\cal A}$. Now, let us define the distribution
\[\tilde {f}(p,g)=f(p)\delta_e(g)\]
where $\delta$ is the Dirac distribution, $g\in G$, and $
e$ is the unit of $G$. $\tilde {f}$ convolutes well with
functions of ${\cal A}$. Indeed, let $a\in {\cal A}$; we have
\[(\tilde {f}*a)(p,g)=\int_G\tilde {f}(p,g_1)a(pg_1,g_1^{
-1}g)dg_1=f(p)a(p,g)\in {\cal A}.\]
(Here we have used the integral notation for the distribution
action on test functions.)

Let ${\cal G}=E\times E$ be the space of the pair groupoid,
where $E$ is, as before, the total space of the frame bundle
over space-time $M$, i.e. ${\cal G}=\{(x,p_1,p_2):p_1,p_2\in
E\;{\rm a}{\rm n}{\rm d}\;\pi_M(p_1)=(\pi_M(p_2))=x\}$, and the
algebra $\tilde {{\cal A}}=C^{\infty}({\cal G},{\bf C} )$ with
convolution as multiplication. The composition law reads
$(x,p_1,p_2)\circ (x,p_2,p_3)=(x,p_1,p_3),\,p_1,p_2,p_3
\in E_x,x\in M$, and the convolution is defined accordingly.
\par
\begin{Proposition}
The mapping $J:\tilde {{\cal A}}\rightarrow {\cal A}$, given by
\[J(f)(\gamma )=f(\pi_M(p),p,pg),\]
for $f\in \tilde {{\cal A}},\,\gamma =(p,g)$, is an isomorphism
of algebras.
\label{(C)}
\end{Proposition}
\par \noindent
{\it Proof} Let $\tilde {f}_1,\tilde {f}_2\in\tilde {{\cal A}}$;
we have
\[(\tilde {f}_1*\tilde {f}_2)(x,p_1,p_2)=\int_{E_x}\tilde {
f}_1(x,p_1,p_3)\tilde {f}_2(x,p_3,p_2)dp_3.\]
We notice that the
fiber $E_x$, for every $x\in M$, is diffeomorphic with the group
$G$, and consequently there is a measure on $E_x$ induced from
the Haar measure on $G$. After making the substitution
$p_3=p_1g_1$, $p_2=p_1g$, we obtain
\[(\tilde {f}_1*\tilde {f}_2)(x,p_1,g_{})=\int_G\tilde {
f}_1(x,p_1,p_1g_1)\tilde {f}_2(x,p_1g_1,p_1g)dg_1\]
 which can be
rewritten as
\[(f_1*f_2)(\gamma )=\int_{\Gamma_{d(\gamma )}}f_1(\gamma_
1)f_2(\gamma_1^{-1}\circ\gamma )d\gamma_1.\;\Box\]

\section{Module of derivations}
Among derivations of the algebra ${\cal A}$ on the groupoid $
\Gamma =E\times G$ we can
distinguish three types: horizontal derivations, verical
derivations, and inner derivations of ${\cal A}$; we denote them
by ${\rm D}{\rm e}{\rm r}_{Hor}{\cal A},\,{\rm D}\,{\rm e}
{\rm r}_{Ver}{\cal A}$, and ${\rm I}{\rm n}_{}{\rm n}{\cal A}$,
respectively. We shall study them in turn.  
\par
\begin{Lemma}
\label{(A)}
Let $\bar {X}\in {\cal X}(E)$ be a right
invariant vector field (on a principal bundle), i.e., $
({\cal R}_g)_{*p}\bar {X}(p)=\bar {X}(pg)$ for every $g
\in G$. Its lifting to $\Gamma$, 
$\bar{\bar {X}}(p,g)=(\iota_g)_{*p}\bar {X}(p)$, where the
inclusion $\iota_g:E\times G$ is defined by $\iota_g(p)
=(p,g)$, is 
a derivation of the algebra $({\cal A},*)$.  
\end{Lemma}
\par \noindent
{\it Proof }
\begin{eqnarray*}
\bar{\bar {X}}(p,g)(f*h)](p,g)&=&[\bar {X}(p)(f*h)](\iota_
g(p))=\\
&&\int_G[\bar {X}(p)f(p,g_1)]h(pg_1,g_1^{-1}g)dg_1+\\
&&\int_Gf(p,g_1)[\bar {(}X)(p)h(pg_1,g_1^{-1}g)]dg_1=\\
&&\int_G[(\bar{\bar {X}}))(p,g_1)f](p,g_1)h(pg_1,g_1^{-
1}g)dg_1+\\
&&\int_Gf(p,g_1)[\bar{\bar {X}}(pg_1,g_1^{-1}g)]h(pg_1,
g_1^{-1}g)dg_1=\\
&&(\bar{\bar {X}}f*h+f*\bar{\bar {X}}h)(p,g).\end{eqnarray*}
We have employed here the right invariance property.  $
\Box$\\

\subsection{Horizontal Derivations}

The group $G$ acts freely and transitively on the fibres of $
E$.  Consequently, the $G$-right-invariant vector fields on $ E$
are determined by their values at a single point of every fiber.
Therefore, they can be identified with the cross sections $
\Sigma =TE/G$ of the bundle. Let us consider the mapping
\[(\pi_M)_{*}:TE\rightarrow TM.\]
Since $(\pi_M)_{*}$ is $G$-invariant, it induces the mapping
\[\pi_M:\Sigma\rightarrow TM.\]
Let us denote $\rho =(\bar{\pi}_M)_{*}$, and 
consider the exact sequence of vector bundles
\[0\rightarrow {\rm k}{\rm e}{\rm r}\rho\stackrel j{\rightarrow}
\Sigma\begin{array}{llll}
\mbox{$\rho$}\\
\mbox{$\rightarrow$}\\
\mbox{$\leftarrow$}\\
\mbox{$\sigma$}\end{array}
TM\rightarrow 0.\]
The mappings $j$ and $\rho$ are homomorphisms of vector bundles,
and $j$ is an inclusion.  The homomorphism of vector bundles
$\sigma :TM\rightarrow\Sigma$ is a connection in the principal
bundle $\pi_M:E\rightarrow M$ if it splits this sequence, i.e.,
if $\rho\circ\sigma ={\rm i}{\rm d}_{TM}$. In our case, such
$\sigma$ always exists although it is not unique.  With the help
of $\sigma$ we lift a vector field $X\in {\cal X}(M)$ from $M$
to $\Sigma$, i.e., $\bar {X}(p)=\sigma
(X(\pi_M(p)),\,\pi_M(p)=x\in M$, and we concider $\bar {X}$ as a
$G$-right-invariant vector field on $E$. Finally, we lift this
field, with the help of the inclusion $\iota_ g$, to the
groupoid $\Gamma$. We thus obtain
\[\bar{\bar {X}}(p,g)=(\iota_g)_{*p}\bar {X}(p)\in {\cal X}
(E\times G)\]
for every $(p,g)\in\Gamma$. Vector fields $\bar{\bar {X}}\in
{\cal X}(\Gamma )$, obtained in this way, inherit from $\sigma$
the right invariance property. Lemma \ref{(A)} 
evidently applies to such vector fields. Moreover, we have the
following proposition.

\begin{Proposition}
Vector fields $\bar{\bar {X}}\in {\cal X}
(\Gamma )$ form a $Z$-submodule of the $Z$-module ${\rm D}
{\rm e}{\rm r}{\cal A}$ of derivations of the algebra $
{\cal A}$.
They will be called {\em horizontal\/} derivations of $
{\cal A}$
and denoted by ${\rm D}{\rm e}{\rm r}$$_{Hor}{\cal A}$.
\end{Proposition}
\par \noindent
{\it Proof} Let $a,b\in {\cal A}$. One readily checks, taking
into account the right invariance of $\bar{\bar {X}}$, that
$f\bar{\bar {X}}(a*b)$, $f\in Z,a,b\in {\cal A}$, satisfies the
Leibniz rule. We shall show that $f\bar{\bar {X}}\in {\rm D}
{\rm e}{\rm r}_{Hor}{\cal A}$. Indeed, let $f_0\in C^{\infty}
(M)$ be such that $f=\pi_M^{*}f_0$, and $X':=f_0X,\,X\in 
{\cal X}(M)$. We have
\[\bar {X}'=\pi_M^{*}f_0\bar {X}=f\bar {X},\]
and by acting on both sides with $\iota_g$ we obtain $\bar{
\bar {X}}'=f\bar{\bar {X}}$. $\Box$

\subsection{Vertical Derivations}
Let us consider all right invariant vertical vector fields on $
E$, i.e., all right invariant vector fields $\bar {X}\in {\cal
X}(E)$ such that $(\pi_M)_{*}(\bar {X})=0$. Such vector fields
lifted to $\Gamma$ are, on the strength of Lemma \ref{(A)},
derivations of the algebra $({\cal A},*)$; we shall call them
{\em vertical\/} derivations of this algebra, i.e.
\[\bar{\bar {X}}(p,g)=(\iota_g)_{*p}\bar {X}(p)\in {\rm D}
{\rm e}{\rm r}_{Ver}{\cal A}\]
for every $g\in G$.
\par
Let us notice that $\bar {X}\in {\cal X}(E)$ can be regarded as
cross sections of the vector bundle ${\rm k}{\rm e}{\rm r}
\rho$ and, as it
can be easily seen, ${\rm D}{\rm e}{\rm r}_{Ver}{\cal A}$ is a $
Z$-submodule of the $Z$-module ${\rm D}{\rm e}{\rm r}{\cal A}$.

\subsection{Inner Derivations}
The set of inner derivations of the algebra ${\cal A}$ is
defined as follows
\[{\rm I}{\rm n}{\rm n}{\cal A}=\{{\rm a}{\rm d}a:a\in 
{\cal A}\}\]
 where $({\rm a}{\rm d}a)(b):=a*b-b*a$.
\par
\begin{Lemma}
\label{(B)} The mapping $\Phi :A\rightarrow 
{\rm I}{\rm n}{\rm n}{\cal A}$, given by $\Phi (a)={\rm a}
{\rm d}a$, is an
isomorphism of Lie algebras (and also of $Z$-moduli).
\end{Lemma}
\par \noindent
{\it Proof} It can be easily seen that
\[[{\rm a}{\rm d}a,{\rm a}{\rm d}b]={\rm a}{\rm d}[a,b]
\in {\cal A},\]
i.e., ${\rm I}{\rm n}{\rm n}{\cal A}$ is a Lie algebra and $
\Phi$ is a Lie
algebra homomorphism. Then we have: $\Phi (a)=\Phi
(b)\Rightarrow [a,c]=[b,c]$, for every $c\in {\cal A}$. Hence
$[a-b,c]=0$ since $a-b\in {\cal Z}({\cal A})=\{0\}$. Therefore,
$a=b$. We also see that
\[\Phi (fa)={\rm a}{\rm d}(fa)=f{\rm a}{\rm d}a=f\Phi (
a)\]
for every $f\in Z$$.\Box$
\par
As we have seen in the proof of this Lemma, the fact that $
{\cal Z}({\cal A})=\{0\}$ plays an important role in the entire
structure.

\subsection{Some Properties of Derivations}
By {\em differential algebra\/} we understand a pair $( {\cal
A},{\rm D}{\rm e}{\rm r}{\cal A})$ where ${\cal A}$ is a not
necessarily commutative algebra and ${\rm D}{\rm e}{\rm r} {\cal
A}$ a (sub)module of its derivations. In the following, we will
base the construction of a noncommutative geometry of the
transformation groupoid $\Gamma$ on the differential algebra
$({\cal A},{\rm D}{\rm e}{\rm r}{\cal A})$ where ${\cal A}$ is,
as above, $C^{\infty}_c(\Gamma ,{\bf C})$, and
\[{\rm D}{\rm e}{\rm r}{\cal A}={\rm D}{\rm e}{\rm r}_{
Hor}{\cal A}\oplus {\rm D}{\rm e}{\rm r}_{Ver}{\cal A}\oplus
{\rm I}{\rm n}{\rm n}{\cal A}.\]

Let $\bar{\bar {X}}_1,\bar{\bar {Y}}_1\in {\rm D}{\rm e} {\rm
r}_{Hor}{\cal A}$, $\bar{\bar {X}}_2,\bar{\bar {Y}}_2\in {\rm
D}{\rm e}{\rm r}_{ Ver}{\cal A}$, ${\rm a}{\rm d}a,{\rm a}{\rm
d}b\in {\rm I} {\rm n}{\rm n}{\cal A}$. We have the following
properties:
\par
\begin{enumerate}
\item
$[\bar{\bar {X}}_1,\bar{\bar {Y}}_1$$]=\overline {\overline {
[X_1,Y_1]}}$. This follows from the fact that $\bar {X} =\sigma
(X)$ which implies that $[\bar {X}_1,\bar {Y}_1 ]=\overline
{[X_1,Y_1]}$.
\par
\item
$[\bar{\bar {X}}_2,\bar{\bar {Y}}_2]=\overline {\overline {
[X_2,Y_2]}}$.
\item
$[{\rm a}{\rm d}a,{\rm a}{\rm d}b]={\rm a}{\rm d}[a,b]\in {\rm
I}{\rm n}{\rm n}{\cal A}$, see proof of Lemma \ref{(B)}.
\item
$[\bar{\bar {X}}_1,\bar{\bar {X}}_2]=0$, since cross sections of
the vector bundle $\Sigma$ form a Lie algebra which splits into
the sum of two Lie subalgebras, and the fields $\bar {X}_1$ and
$\bar {X}_2$ belong to different subalgebras.
\item
$[\bar{\bar {X}}_1,{\rm a}{\rm d}a]={\rm a}{\rm d}\bar{
\bar {X}}_1(a)$, by simple computations.
\item
$[\bar{\bar {X}}_2,{\rm a}{\rm d}a]={\rm a}{\rm d}\bar{
\bar {X}}_2(a)$, by simple computations.
\end{enumerate}

\section{Geometry of ${\rm D}{\rm e}{\rm r}_{Ver}{\cal A}$ and $
{\rm I}{\rm n}{\rm n}{\cal A}$}
Because of the decomposition of the $Z$-module ${\rm D}
{\rm e}{\rm r}{\cal A}$ into three parts, the metric on $
{\rm D}{\rm e}{\rm r}{\cal A}$
\[{\cal G}:{\rm D}{\rm e}{\rm r}{\cal A}\times {\rm D}{\rm e}
{\rm r}{\cal A}\rightarrow Z\]
 also decomposes into three parts.
If $u=u_1+u_2+u_3$, and $u_1\in {\rm D}{\rm e}{\rm r}_{
Hor}{\cal A}$, $u_2\in {\rm D}{\rm e}{\rm r}_{Ver}{\cal A}$,
$u_3\in {\rm I}{\rm n}{\rm n}{\cal A}$, and analogopusly for
$v=v_1+v_2+v_3$, then
\[{\cal G}(u,v)=\bar {g}(u_1,v_1)+\bar {k}(u_2,v_2)+h(u_
3,v_3)\]
 where $\bar {g}:{\rm D}{\rm e}{\rm r}_{Hor}{\cal A}\times 
{\rm D}{\rm e}{\rm r}_{Hor}{\cal A}\rightarrow Z$ is
evidently the lifting of the metric $g:{\cal X}(M)\times 
{\cal X}(M)\rightarrow C^{\infty}(M)$ on space-time $M$, i.e.,

\[\bar {g}(u_1,v_1)=\pi_M^{*}(g(x,y))\]
where $x,y\in {\cal X}(M)$.  We assume that the metrics $
\bar {k}:{\rm D}{\rm e}{\rm r}_{Ver}{\cal A}\times {\rm D}
{\rm e}{\rm r}_{Ver}{\cal A}\rightarrow Z$ and $h:{\rm I}
{\rm n}{\rm n}{\cal A}\times {\rm I}{\rm n}{\rm n}{\cal A}
\rightarrow Z$ are of the Killing type.  Their form will be
determined below.
\par\ 
The preconnection is given by the Koszul formula
\begin{eqnarray}
(\nabla^{*}_uv)w=\frac 12[u({\cal G}(v,w))+v({\cal G}(u
,w))-w({\cal G}(u,v))\\
+{\cal G}(w,[u,v])+{\cal G}(v,[w,u])-{\cal G}(u,[v,w])]
.\nonumber\label{Koszul}\end{eqnarray}
 
Let us now consider a more general situation which will later be
specified to that in our model.  Let $({\cal A} ,*)$ be an
algebra over {\bf C,} $Z={\cal Z}({\cal A})$ its center, and
${\cal V}$ a $Z$-module of derivations of the algebra ${\cal
A}$.  What follows is also valid if ${\cal Z} ({\cal A})=\{0\}$
and $Z=\pi_M^{*}(C^{\infty}(M))$ as in our model.  We further
assume that elements of $Z$ play the role of constants for
derivations of ${\cal V}$, i.e.,  ${\cal V} (Z)=\{v(f) = 0 :v\in
{\cal V},f\in Z\}$.  Let us consider a metric $g:{\cal
V}\times {\cal V}\rightarrow Z$; we assume the $ Z$-2-linearity
and symmetry of $g$, but not necessarily its nondegeneracy.  Let
us denote ${\cal V}^{*}={\rm H}{\rm o} {\rm m}({\cal V},Z)$, and
$u^{*}=g(u,\cdot )=\Phi_g(u)$ is a one-form corresponding to the
derivation $v\in {\cal V}.$
\par
The symmetric two-form $g$ determines the preconnection $
\nabla^{*}:{\cal V}\times {\cal V}\rightarrow {\cal V}^{
*}$ by the Koszul formula (\ref{Koszul}) (with ${\cal G}$
replaced by $g$). Since, by assumption, ${\cal V}(Z)=\{
0$$\}$,
one has $\nabla^{*}_u(fv)=f\nabla^{*}_uv$, and $\nabla^{
*}$ is a
$Z$-2-linear mapping, i.e.  a tensor of (2,1) type. Moreover,
from the Koszul formula it follows (even if $g$ is degenerate)
that
\[w(g(u,v))=(\nabla^{*}_wu)(v)+(\nabla^{*}_wv)(u)\]
In the Koszul formula the first three terms vanish, and if we
assume that
\begin{equation}g(v,[w,u])=g(u,[v,w]),\label{eqA}\end{equation}
which --- as we shall see below --- is valid in our case, we
obtain an interesting result
\[(\nabla^{*}_uv)(w)=g(\frac 12[u,v],w)\]
showing that there is a strict depedence between the
(pre)connection and the metric. We should only look for a
mapping $\nabla :{\cal V}\times {\cal V}\rightarrow {\cal V}$
that would be $g$-consistent with $\nabla^{*}:{\cal V}\times
{\cal V}\rightarrow {\cal V}^{*}$, i.e. satisfying the condition
\[(\nabla^{*}_uv)(w)=g(\nabla_uv,w)\]
for every $u,v,w\in {\cal V}$. By comparison with (\ref{eqA}),
it immediately follows that
\[\nabla_uv=\frac 12[u,v]\]
for every $u,v\in {\cal V}$. Moreover, for a nondegenerate
tensor $g$ the mapping $\nabla$, $g$-consistent with
$\nabla^{*}$, is unique. Indeed, since from
\[(\nabla^{*}_uv)(w)=g(\stackrel 1{\nabla}_uv,w)=g(\stackrel 
2{\nabla}_uv,w),\]
for every $u,v\in {\cal V}$, it follows that
\[\stackrel 1{\nabla}_uv=\stackrel 2{\nabla}_uv.\]
\par
It turns out that if $g$ is nondegenerate, it has all properties
required for connection. Let us check, for instance,
\begin{eqnarray*}
(\nabla^{*}_{u_1+u_2}v)(w)&=&(\nabla^{*}_{u_1}v)(w)+(\nabla^{
*}_{u_2}v)(w)\\
&=&g(\nabla_{u_1}v+\nabla_{u_2}v,w)\\
&=&g(\nabla_{u_1+u_2}-\nabla_{u_1}v-\nabla_{u_2}v,w)=0,\end{eqnarray*}
and, from the nondegeneracy of $g$, one has
\[\nabla_{u_1+u_2}v=\nabla_{u_1}v+\nabla_{u_1}v.\]
\par
Therefore, we have proved the following proposition
\par
\begin{Proposition}
Let ${\cal V}$ be a $Z$-module of derivations of an algebra
$({\cal A},*)$ such that ${\cal V}(Z)=\{0
\}$. For every symmetric nondegenerate tensor $g:{\cal V}
\times {\cal V}\rightarrow Z$, there exists one and only one connection
$g$-consistent with the preconnection $\nabla^{*}$. It is given
by
\[\nabla_uv=\frac 12[u,v].\;\Box\]
\end{Proposition}
\par
In the following, we shall assume the metric of the Killing form
\[g(u,v)={\rm T}{\rm r}(u\circ v).\]
It satisfies the $g$-consistency condition. Indeed, from the
trace definition we have
\[{\rm T}{\rm r}[w\circ u,v\circ u]={\rm T}{\rm r}([w,u
]\circ v)+{\rm T}{\rm r}([w,v]\circ u)=0.\]
\par 
We now return to our model, and assume the above kind of metric
for both ${\rm D}{\rm e}{\rm r}_{Ver}{\cal A}$ and ${\rm I} {\rm
n}{\rm n}{\cal A}$, but in both these cases the trace should be
defined differently.
\par
We first define the metric for ${\rm D}{\rm e}{\rm r}_{
Ver}{\cal A}$. We assume that $G$ is a semisimple group. In this
case, the Killing form reads
\[{\cal B}(V,W)={\rm T}{\rm r}({\rm a}{\rm d}V\circ {\rm a}
{\rm d}W),\]
for $V,W\in\underline g$, where $\underline g$ is the Lie
algebra of the group $G$, and  ${\cal B}$ is nondegenerate.  The
tangent space to any fiber $E_x,\,x\in M$, is isomorphic to
$\underline g$.  Therefore, the metric $\bar {k}:\mbox{${\rm
D}{\rm e} {\rm r}_{Ver}{\cal A}$$\times\mbox{${\rm D}{\rm e}{\rm
r}_{ Ver}{\cal A}$$\rightarrow Z$}$}$ is given by
\[\bar {k}(\bar{\bar {X}},\bar{\bar {Y}})={\cal B}(X(\pi_
M(p),Y(\pi_M(p)).\]
\par
To define the metric for ${\rm I}{\rm n}{\rm n}{\cal A}$, let us
first define the trace for the algebra $\tilde {{\cal A}}$
(which, by Proposition \ref{(C)}, is isomorphic to the algebra
${\cal A}$), $\mbox{T}{\rm r}:{\cal A}\rightarrow Z$, by
\[({\rm T}{\rm r}a)(x)=\int_Ga(x,g,g)dg.\]
It has the following properties:
(i) ${\rm T}{\rm r}(a+b)={\rm T}{\rm r}a+{\rm T}{\rm r}
b,$
(ii) ${\rm T}{\rm r}(fa)=f{\rm T}{\rm r}a$,
(iii) ${\rm T}{\rm r}(a*b)={\rm T}{\rm r}(b*a)$,
for $a,b\in {\cal A},\,f\in Z$. From the last property it
follows that
\[{\rm T}{\rm r}([a,b])=0,\]
and, of course, ${\rm T}{\rm r}\circ {\rm a}{\rm d}a=0$.
\par
Let us now turn to the submodule Inn${\cal A}$. We should notice
that, on the strength of Lemma \ref{(B)}, we also have the
connection $\tilde{\nabla }:{\cal A}\times {\cal A}\rightarrow
{\cal A}$ on $ {\cal A}$ given by
\[\tilde{\nabla}_ab=\frac 12[a,b].\]
\par\ 
We define the metric $h:{\rm I}{\rm n}{\rm n}{\cal A}\times {\rm
I}{\rm n}{\rm n}{\cal A}\rightarrow Z$ by
\[h({\rm a}{\rm d}a,{\rm a}{\rm d}b)={\rm T}{\rm r}(a*b
),\]
 and the corresponding connection is
\[\nabla_{{\rm a}{\rm d}a}{\rm a}{\rm d}b=\frac 12[{\rm a}
{\rm d}a,{\rm a}{\rm d}b].\]
\par
We shall show that the metric $h$ is nondegenerate. Indeed, let
us assume that
\[{\rm T}{\rm r}(a*b)=\int_G\int_Ga(x,g_{1,}g_2)b(x,g_2
,g_1)dg_2dg_1=0.\]
If $a\neq 0$, then the support of this function is not of the
measure zero, and by choosing the function
$b(x,g_2,g_1)=a(x,g_1,g_2)$, we obtain
\[\int_G\int_Ga^2(x,g_1,g_2)dg_2dg_1\neq 0.\]
We conclude that if the metric is of the trace type (either $
{\cal B}$ or Tr), the formula (\ref{eqA}) is valid (for $
g=\bar {k}$, or $g=h$).

\section{Curvature} 
Let us introduce the following abbreviations:
\[V_1={\rm D}{\rm e}{\rm r}_{Hor}{\cal A},\;\;V_2={\rm D}
{\rm e}{\rm r}_{Ver}{\cal A},\;\;V_3={\rm I}{\rm n}{\rm n}
{\cal A}.\]
 
We continue to develop the geometry for $V_i,\,i=1,2,3$. The
curvature is
\[\stackrel iR:V_i\times V_i\times V_i\rightarrow V_i,\]
\[\stackrel iR(u,v)w=\stackrel i{\nabla}_u\stackrel i{\nabla}_
vw-\stackrel i{\nabla}_v\stackrel i{\nabla}_uw-\stackrel 
i{\nabla}_{[u,v]}w.\]
\par
If $j=2,3$, we have
\begin{eqnarray*}
\stackrel jR(u,v)w&=&\frac 12[u,\frac 12[v,w]]\\
&-&\frac 12[v,\frac 12[u,w]-\frac 12[[u,v],w]\\
&=&-\frac 14[[u,v],w]].\end{eqnarray*}
 
Here we have made use of the Jacobi identity.
\par
For every endomorphism $T:V_i\rightarrow V_i,\,i=1,2$, there
exists ${\rm T}{\rm r}T\in Z$ satisfying the usual trace
conditions.  We thus can define
\[\stackrel i{R^m}_{uwm}:V_i\rightarrow V_i,\]
\[\stackrel iR_{uw}(v)=\stackrel iR(u,v)w,\]
and
\[\stackrel i{{\bf r}{\bf i}{\bf c}}:V_i\times V_i\rightarrow
Z\]
\[\stackrel i{{\bf r}{\bf i}{\bf c}}(u,v)={\rm T}{\rm r}\stackrel 
iR_{uw}.\]
\par
We also define the {\em adjoint Ricci operator}
\begin{equation}\stackrel i{{\bf r}{\bf i}{\bf c}}(u,w)
=\stackrel i{{\cal G}}(\stackrel i{{\cal
R}}(u),w)\label{Ricci}\end{equation} where we have introduced
the notation: $\bar {g}=\stackrel 1{{\cal G}},\,\bar
{k}=\stackrel 2{{\cal G}}$. If the metric $\stackrel i{{\cal
G}}$ is nondegenerate, there exists the unique $\stackrel
i{{\cal R}}$ satisfying eq. (\ref{Ricci}) for every $v\in V_i$.
\par
The {\em curvature scalar\/} is
\[\stackrel ir={\rm T}{\rm r}\stackrel i{{\cal R}}\in Z
.\]
\par
In the module $V_2$ there exists the usual trace operator which,
in the local basis, can be written as the trace of the operator
matrix. Therefore,
\[\stackrel 2R_{uv}(w)=\frac 14[w,[u,v]]=\frac 14({\rm a}
{\rm d}w\circ {\rm a}{\rm d}u)(v),\]
 
and we have
\[\stackrel 2R_{uw}=\frac 14({\rm a}{\rm d}w\circ {\rm a}
{\rm d}u)\]
 
for every $u,w\in V_2$,
and
\[{\rm T}{\rm r}\stackrel 2R_{uw}=\frac 14{\rm T}{\rm r}
({\rm a}{\rm d}w\circ {\rm a}{\rm d}u).\]
 
Hence,
\begin{equation}\label{Ric2}\stackrel 2{{\bf r}{\bf i}{\bf c}}
(u,w)=\frac 14\bar {k}(u,w)\end{equation}
for every $u,w\in V_2$, which can be regarded as a generalized
Killing form. By analogy, we postulate
\begin{equation}\label{Ric3}\stackrel 3{{\bf r}{\bf i}{\bf c}}
(u,w)=\alpha h(u,v)\end{equation}
for every $u,w\in V_3$.
\par
The ``Ricci scalar'' can be determined from the generalized
Einstein equation
\[\stackrel 3{{\bf r}{\bf i}{\bf c}}(u,v)-\frac 12rh(u,
v)=0\]
or
\[\alpha h(u,v)-\frac 12rh(u,v)=0.\]
Hence we obtain
\[(\alpha -\frac 12r)h(u,v)=0,\]
and
\[\alpha =\frac 12r.\]
We can symbolically regard $r\in Z$ as a trace of the Ricci
operator $ {\cal R}$. The Ricci 2-form is thus proportional to
the metric tensor $ h$, and the proportionality coefficient (up
to factor 2) is a counterpart of the Ricci curvature scalar.
\par
A counterpart of eq. (\ref{Ricci}) for $V_3$ is
\begin{equation}\stackrel 3{{\bf r}{\bf i}{\bf c}}(u,v)
=h(\stackrel {3}{\cal R}(u),v).\label{3Ricci}\end{equation}
Hence
\[\alpha h(u,v)=h(\stackrel 3{{\cal R}}(u),v),\]
or
\[h(\alpha u,v)=h(\stackrel 3{{\cal R}}(u),v),\]
and finally,
\[\stackrel 3{{\cal R}}(u)=\alpha u.\]
\par
Let us notice that for a commutative algebra we have $\alpha 
=0$, and the sector 
corresponding to $V_3$ vanishes. Therefore, the coefficient $
\alpha$ could be regarded 
as a ``measure'' of noncommutativity.
\par
This concludes the construction of the noncommutative groupoid
geometry.  The transition from this geometry to the usual
space-time geometry can be done by the following ``averaging''
procedure. If $a\in {\cal A}$ then we have the isomorphism
$a(p,g)=\tilde {a}(x,g_1,g_2)$, and we define
\[\langle \tilde {a}\rangle (x)=\int_G\tilde a(x,g,g)dg.\]
It is clear that $\langle \tilde a\rangle\in C^{\infty}_c(M)$,
and from the algebra $C^{\infty }_c(M)$ one can reconstruct the
usual space-time geometry together with the usual Einstein
equations \cite{Geroch72}.

\section{Generalized Einstein's equation} 
We have all geometric quantities necessary to write the
counterpart of Einstein's equation on the groupoid $\Gamma$. We
stipulate that in the noncommutative regime at the fundamental
level, there is only a ``pure noncommutative geometry'', and all
necessary ``matter terms'' will somehow emerge out of it. We
thus assume that there is no counterpart of the energy-momentum
tensor and, consequently, the generalized Einstein's equation is
of the form
\begin{equation}{\cal R}-\frac 12r\,{\rm i}{\rm
d}_V=0\label{Einstein}\end{equation} 
where ${\cal R}$ is the Ricci operator defined by eq.
(\ref{Ricci}) (superscipt $ i=1,2$ is omitted but presupposed),
and $r={\rm T}{\rm r}{\cal R}$.
\par
It is clear that eq. (\ref{Einstein}), for $V=V_1$, is a
``lifting'' of the usual Einstein equation on space-time $ M$ to
the groupoid $\Gamma$, and every $\bar {g}$ that solves this
equation on $ M$ solves also eq. (\ref{Einstein}).
\par
Let us now consider the case $V=V_2$. By comparing eq.
(\ref{Ricci}) with eq. (\ref{Ric2}) and noticing that $\frac
14\bar {k}(u,v)=\bar {k}(\frac 14u,v)$, one obtains
\[\stackrel 2{{\cal R}}=\frac 14{\rm i}{\rm d}_{V_2}.\]
\par
Similarly, for $V=V_3$, by taking into account eq. (\ref{Ric3})
and comparing it with eq. (\ref{Ric2}), we obtain
\[\stackrel 3{{\cal R}}=\alpha\,{\rm i}{\rm d}_{V_3}.\]
\par
Let us consider the ${\cal G}$-orthogonal sum $V=V_1\oplus
V_2\oplus V_3$. For the Ricci operator ${\cal R}:V\rightarrow V$
we have ${\cal R}(V_i)\subseteq V_i,\,i=1,2,3$, and $ {\cal
R}|V_i=\stackrel i{{\cal R}}$. This leads to the eigenvalue
equation
\[{\cal R}(u)=\lambda u\]
for $u\in V$. This eigenvalue problem has the following
solutions:
\begin{enumerate}
\item
$\lambda_1=\frac 12r$ where $r$ is the Ricci scalar curvature
for the metric tensor $
\bar {g}$. We thus have the equation
\[{\cal R}(u)-\frac 12ru=0\]
for $u\in V_1$, and each such $u$ satisfies this equation. It
can be easily checked that this equation reduces to the equation
$ {\cal R}=0$ on space-time $M$.
\item
$\lambda_2=\frac 14$ which leads to the equation
\[{\cal R}(u)-\frac 14u=0\]
for $u\in V_2$.
\item
$\lambda_3=\alpha$ leading to the equation
\[{\cal R}(u)-\alpha u=0\]
for $u\in V_3$. In the commutative case $\alpha = 0$ and we
obtain ${\cal R}(u) = 0$ (on the groupoid $\Gamma $).
\end{enumerate}

\section{Quantum sector}
The quantum sector of our model is obtained by the regular
representation of the groupoid algebra ${\cal A}$ in a Hilbert
space ${\cal H}^p=L^2(\Gamma^p)$, $p\in E$,
\[\pi_p:{\cal A}\rightarrow {\cal B}({\cal H}^p),\]
where ${\cal B}({\cal H}^p)$ denotes the algebra of bounded
operators on ${\cal H}^p$, given by
\[(\pi_p(a)\psi )(\gamma )=\int_{\Gamma_{d(\gamma )}}a(
\gamma_1)\psi (\gamma_1^{-1}\circ\gamma )d\gamma_1\]
where $a\in {\cal A},\,\psi\in {\cal H}^p,\,\gamma ,\,\gamma_
1\in\Gamma$. Let us notice that the Haar measure on the group
$G$, transferred to the fibres of $\Gamma$, forms a Haar system
on $\Gamma$.
\par
We shall show that every element $a\in {\cal A}$ generates a
random operator $r_a$ on $({\cal H}^p)_{p\in E}$. By a {\em
random operator} $r$ we mean a family of operators $ (r_p)_{p\in
E}$, i.e., a function
\[r: E \rightarrow\bigsqcup_{p\in E}{\cal B}({\cal H}^p)\]
such that
\begin{enumerate}
\item
the function $r$ is measurable in the following sense: if $
\xi^{}_p,\eta_p\in {\cal H}^p$ then the function
\[E\ni p\mapsto (r(p)\xi_p,\eta_p)\in {\bf C}\]
is measurable with respect to the manifold measure on $
E$;
\item
the function $r$ is bounded with respect to the norm:
\[||r||={\rm e}{\rm s}{\rm s}\,{\rm s}{\rm u}{\rm p}||r
(p)||\]
where ${\rm e}{\rm s}{\rm s}\,{\rm s}{\rm u}{\rm p}$ means the
``supremum modulo zero measure sets''.
\end{enumerate}
\par
Random operator $r$ acts, in fact, on cross sections of the
Hilbert bundle ${\cal H}=\bigsqcup_{p\in E}{\cal H}^p$.
\par
It can be easily seen that the family of operators
$r_a=(\pi_p(a))_{p\in E}$ is a random operator. Indeed, if
$\xi_p,\,\eta_p\in L^2(\Gamma^p)$ then we have the scalar
product
\[(\int_{\Gamma^p }\pi_p(a)\xi_p,\eta_p)=\int_{ \Gamma^p
}(\int_{\Gamma_{d(\gamma )}}a(\gamma_
1)\xi_p(\gamma_1^{-1}\circ\gamma )d\gamma_1)\overline {
\eta_p(\gamma )}d\gamma ,\]
and the Haar measure is transferred
from $G$ to $\Gamma^ p$ for each $p\in E$. Therefore, condition
(1) is satisfied.
\par
To check condition (2) let us introduce the isomorphisms of
Hilbert spaces $I_p: L^2(G) \rightarrow {\cal H}^p$ given by the
formula 
\[
(I_p\psi)(pg^{-1},g)=\psi(g) 
\] 
for $\psi \in L^2(G)$.
Let us consider the operators $\tilde {\pi}_p (a)= I_p^{-1}\circ
\pi_p (a) \circ I_p$. It is clear that $||\pi_p(a)|| =
||\tilde{\pi}_p(a)||$. Let us also notice that
\[
\tilde {\pi}_{pg}(a)=R_{g^{-1}}\circ\tilde{\pi}_p(a)\circ R_g 
\] 
(where
$R_g$ denotes the right translation operator in the space
$L^2(G)$), which entails the (unitary) invariance of the norm
\[||\pi_{pg}(a)||=||\pi_p(a)||.\]
Hence, the norm $||\pi_p(a)||$ depends only on $x=\pi_ M(p)\in
M;$ therefore, the function $x\mapsto ||\pi_p(a)||$ is well
defined, compactly supported and continuous (in its dependence
on $x$) on $M$.
\par
Let ${\cal M}$ denote the set of all equivalence classes
(modulo equality almost everywhere) of random operators
$r_a,\,a\in {\cal A}$. It forms a von Neumann algebra; we shall
call it the {\em algebra of random operators of the groupoid}
$\Gamma$, or simply the {\em von Neumann algebra of the groupoid}
$\Gamma$.  We shall show that ${\cal M}$ is a semifinite algebra
and, consequently, that it admits a ``modular evolution'', just
like in the model with a finite group $ G$ \cite{Obs}.  To this
end, let us first recall some important concepts.
\par
A von Neumann algebra ${\cal M}$ is {\em semifinite\/} if there
exists a faithful, normal, semifinite weight $\varphi$ on
${\cal M}$ which is a trace.
\begin{itemize}
\item
A linear functional $\varphi :{\cal M}\rightarrow {\bf C}$ is a
{\em state\/} on ${\cal M}$ if $\varphi (r)\geq 0$ for every $r\in
{\cal M}_{+}$, where ${\cal M}_{+}=\{x\cdot x^{*}:x\in {\cal M}\}$
is the subset of positive elements of ${\cal M}$ and $\varphi
(1)=1$.
\item\
A functional $\varphi :{\cal M}_{+}\rightarrow [0,\infty ]$ is a
{\em weight \/} if $\varphi$ is additive, i.e., $\varphi
(x+y)=\varphi (x)+\varphi (y)$, and positively homogeneous,
i.e., $\varphi (\lambda x)=$$\lambda\varphi (x)$, for ${\bf R}
\ni\lambda\geq 0$, $x,y\in {\cal M}$.  We additionally assume: $
\lambda +\infty =\infty$, $\lambda\cdot\infty =\infty$ if
$\lambda\neq 0$, and $\lambda \cdot\infty =0$ if $\lambda =0.$
Let us notice that every state defines a weight.
\item
A weight $\varphi$ is {\em faithful\/}  if for $r\in {\cal M}_{
+}$ one has: $\varphi (r)=0\Rightarrow r=0$.
\item
The sufficient and necessary condition for a weight $\varphi$ to
be {\em normal\/} is: $\varphi (x)=\sum_i\omega_i$ for a
family $\{\omega_i\}$ of normal states, i.e., $\omega (r)={\rm
T}{\rm r}(\rho r),\,{\rm T}{\rm r}(\rho
\}=1$ \cite{Sunder}.
\item
Let us define: $D_{\varphi}:=\{x\in {\cal M}_{+}:\varphi
(x)<\infty \}$ and ${\cal M}_{\varphi}:= {\rm S}{\rm p} {\rm
a}{\rm n}_{{\bf C}}(D_{\varphi})$, i.e. ${\cal M}_{\varphi}$ is
the space of {\bf C}-linear combinations of elements of
$D_{\varphi}$. A weight $ \varphi$ is {\em semifinte\/} if ${\cal
M}_{\varphi}$ is $\sigma$-weakly dense in ${\cal M}$ \cite[p.
56]{Sunder}.
\item
A weight $\varphi$ is a {\em trace\/} if $\varphi (r^{*}
\cdot r)=\varphi (r\cdot r^{*})$, for every $r\in {\cal M}$.
\end{itemize}
\par
\begin{Proposition}
The von Neumann algebra ${\cal M}$ of the
groupoid $ \Gamma$ is semifinite.
\end{Proposition}
\par \noindent
{\em Proof\/} We can consider the von
Neumann algebra ${\cal M}$ as an algebra of bounded operators on
the Hilbert space $H=L^2_G(E,{\cal H})$ of $G$-covariant
square-integrable sections of the bundle ${\cal H}$. $H$ is
isomorphic to $L^2_G(E,L^2(G))\simeq L^2(M\times G)$. The latter
space is separable ($M\times G$ is a locally compact manifold).
We choose the Hilbert basis $\{\psi_k\}_{k=1}^{\infty}$ in $ H$,
and define the weight $\varphi :{\cal M}_{+}\rightarrow
[0,\infty ]$ by
\[\varphi (r)=\sum_{k=1}^{\infty}(r\psi_k,\psi_k).\]
\par
This weight is clearly faithful and trace. It is also normal since
$\varphi_i=\sum_{i=1}^{\infty}\omega_i$ where $\omega_i$ is given
by $ \omega_i(r)={\rm T}{\rm r}(r\rho_i)$ with $\rho_i$
being the projection onto the basis vector $\psi_i\in H$.
\par
To show that $\varphi$ is semifinite, let us notice that we have
the net of finite-dimensional projections $P_{\alpha}$ such
that $\varphi (P_{\alpha})<\infty$ and $\lim P_{\alpha}={\bf
1}${\bf ,} in strong topology, i.e., for every $h\in H$ one has
$P_{\alpha} h=h$. And this is the necessary and sufficient
condition for $\varphi$ to be semifinte \cite[p. 57]{Sunder}.
$\Box$
\par
The fact that the von Neumann algebra ${\cal M}$ is semifinite
ensures that it admits a modular group of automorphisms
\cite[Chapt. 2]{Sunder}. In our case, this group can be defined
for a state (the assumption that $\varphi $ is a weight was
necessary only to prove that ${\cal M}$ is semifinite). Let us
consider a functional of the form
\[\varphi (r)=\int_E{\rm tr}(\hat{\rho }(p)r(p))d\mu_E(p)\]
where $\hat{\rho }(p)$ is a positive operator of trace class
in ${\cal B} ({\cal H}^p)$, for every $p\in E$. Let $\{e_i\}$ be a
basis in ${\cal H}^p$ such that $\hat{\rho }(p)e_i=\lambda_
i(p)e_i,\,\lambda_i>0$. We also postulate 
\par
\[\sum_{i=0}^{\infty}\lambda_i(p)=\lambda (p)<\infty\]
for almost every $p\in E$, and $\lambda(\cdot )\in L^1(E )$ with
\[\int_E\lambda (p)d\mu_E(p)=1.\]
With these conditions the funcional $\varphi$ is a
state, and it satisfies all conditions of the Tomita-Takesaki
theorem. We thus can write the state dependent evolution of random
operators $r\in {\cal M}$ as
\[\sigma_s^{\varphi}(r(p))=e^{isH^{\varphi}(p)}r(p)e^{-i
sH^{\varphi}(p)}\]
where $H(p)={\rm L}{\rm o}{\rm g}\hat{\rho }(p)$ and ${\rm L}
{\rm o}{\rm g}\hat{\rho }(p)e_i=({\rm l}{\rm o}{\rm g}\lambda_
i(p))e_i$. After differentiating the above equation it can be
rewritten as
\begin{equation}
\frac{d}{ds}|_{s=0} r_a(p,s) = i[H^{\varphi }(p), r_a(p)].
\label{Evolution}
\end{equation}
This is a generalization of the Heisenberg equation of the
standard quantum mechanics with the only difference that now the
dynamics depends on the state $\varphi $. The fact we have just
proved that the von Neumann algebra ${\cal M}$ is semifinite,
has serious consequences in this respect.  The Dixmier-Takesaki
theorem \cite[p. 470]{Connes} states that if ${\cal M}$ is
semifinite then every state-dependent evolution is inner
equivalent to the trivial one, i.e.,
\[ U_s \sigma_s^{\varphi}(r(p)) U_s^* = r(p) \] 
for every $s \in \bf{R}$, where $U_s$ is an unitary element of
${\cal M}$. This means that the state-independent evolution,
obtained by the Connes-Nicodym-Radon construction \cite[p.
74]{Sunder}, is trivial. To overcome this difficulty we should
assume that the group $G$ is a locally compact non-unimodular
group. We will return to this problem in a forthcoming paper.
\par
However, a dependence of dynamics on a state need not be a
drawback when we are dealing with the Planck level. The theory
of von Neumann algebras can be regarded as a noncommutative
counterpart of the measure theory. In the commutative case there
is only one interesting measure (the Lebesgue measure), whereas
in the noncommutative case there is a great variety of measures
(see, for instance \cite{Voiculescu}). Each pair $({\cal M},
\varphi )$, where ${\cal M}$ is a von Neumann algebra and
$\varphi $ a state on ${\cal M}$ (usually assumed to be faithful
and normal), is both a dynamic object and a probabilistic
object. In this context, the fact that there are as many
$\varphi $-dependent dynamics as are generalized probabilistic
measures $\varphi $ seems quite natural.

\section{Transition to quantum mechanics}
Let $a^{*}(\gamma )=\overline {a(\gamma^{-1})}$, $a\in {\cal
A},\,\gamma\in\Gamma$, and let us denote by ${\cal A}_ H$ the
subset of all Hermitian elements of ${\cal A}$, i.e. such that
$a^*=a$. If $a\in {\cal A}_H$ then $ \pi^p(a)\in ({\cal
B}({\cal H}^p))_H$ since $\pi_p$ is a $*-$representation of
the algebra ${\cal A}$. In the following we shall consider the
random operators of the form $ r_a(p)=\pi_p(a)$. Such operator is
Hermitian if $(r_a(p)\psi_{},\varphi )=(\psi ,r_a(p)\varphi )$.
Moreover, it is a compact operator for $a\in {\cal A}$, since $a$
has compact support. On the strength of the spectral theorem
for Hermitian compact operators in a separable Hilbert
space, there exists in ${\cal H}^p$ an orthonormal countable
Hilbert basis of eigenvectors $ \{\psi_i\}_{i\in I}$ of the
Hermitian operator $r_a(p)$. We can write its eigenvalue equation
as
\[r_a(p)\psi_i(p)=\lambda_i(p)\psi_i(p).\]
Let us notice that this equation is valid ``for every $p\in E$''
which refelcts the fact that the random operator $ $$ r_a$ is a
family of functions indexed by $p\in E$. Therefore, with respect
to a random operator it is meanigful to speak only about its
eigenfunction $\lambda_i:E\rightarrow {\bf R}$ (not about its
eigenvalue).  However, every concrete measurement is always
performed in a given local frame $p\in E$, and when such a
measurement has been done the eigenfunction $\lambda_i$
collapses to the eigenvalue $\lambda_i(p)$. Let us observe that
from the perspective of the local measurement it looks as if the
measurement result were a random effect, but in fact it is but a
value of a well determined function $\lambda_ i(p)$ at a given
$p$. Its ``randomness'' comes from a subtler source, namely from
the fact that $ r_a$ is a random operator.  This is our model's
version of the so often discussed ``collapse of the wave
function''.
\par
Let us also notice that avery act of measurement, performed at
$p$, singles out the isomorphism $I_p ^{-1}: {\cal H}^p
\rightarrow L^2(G)$ which reproduces the usual quantum mechanics
(on $G$). For instance, to obtain the quantum evolution for
$a\in {\cal A}$, we apply $ I^{-1}_p$ to the left hand side of
equation (\ref{Evolution}), and
$I_p$ to its right hand side. In this way, we obtain
\[\frac d{dt}|_{t=o}\tilde{\pi }(a(t))=i[\tilde {H}^{\varphi}
,\tilde{\pi }(a(t))]\]
where $\tilde {\pi }(a)= I_p^{-1}\circ \pi_p \circ I_p$ and
$\tilde {H}^{\varphi }=I_p^{-1}\circ H_p^{\varphi}\circ I_p$, we
have also put $s = t$. This is the Heisenberg equation of the
standard quantum mechanics with the only difference that it
depends on the state $\varphi$. In more realistic models, to
which the Connes-Nikodym-Radon construction applies, even this
difference disappears (see remarks at the end of the preceding
section).
\par
The above results seem to be important as far as the
interpretation of quantum mechanics is concerned. Its
peculiarities are largely due to the fact that it is but a part
of a larger structure, out of which it is cut off by every act
of measurement. When such an act is performed the larger
structure ``collapses'' to its substructure known as quantum
mechanics. Quantum mechanics turns out to be but a theory of
making measurements within our model.

\end{document}